\documentclass[fleqn,twoside]{article}
\usepackage{espcrc2}

\usepackage{epsfig}

\usepackage[figuresright]{rotating}

\newcommand{\AmS}{{\protect\the\textfont2
  A\kern-.1667em\lower.5ex\hbox{M}\kern-.125emS}}

\title{A New Neutrino Cross Section Data Resource}

\author{M.R. Whalley\address[IPPP]{HEPDATA Database Group and Institute for Particle Physics
                                   Phenomenology,\\ 
        Physics Department, University of Durham, 
        Durham City, DH1 3LE, UK}
	\thanks{work done in collaboration with: \vspace*{0.05cm}\hfill\break 
	C Andreopoulos - CCLRC, Didcot, OX110QX UK \hfill\break
	H Gallagher - Tufts U. Medford MA 023155 USA \hfill\break 
	E Hawker - U. of Cincinnati, Cincinnati OH 45221 USA \hfill\break
	M Sakuda - KEK Ibaraki-ken 305-0801 Japan\hfill\break
	G Zeller - Columbia U. New York NY 10027 USA 
	}
}
       
\begin{document}

\begin{abstract}
We describe a new web based data resource being developed to provide access to 
accurate and validated  cross sections of low energy neutrino and antineutrino interactions.
The proposed content of this database are outlined which cover total and differential cross
from inclusive,
quasi-elastic and exclusive pion production processes from charged and neutral current interactions.
Efforts to obtain these data, which come mainly from old bubble chamber experiments, are 
described as well as the implementation of an embryonic web site to make the resource generally accessible.

\vspace{1pc}
\end{abstract}  

\maketitle

\section{PREAMBLE}

This talk will introduce a 
new and developing resource, namely a database of cross sections of low energy neutrino interactions. 
This is being constructed
by the author in collaboration with those listed below.

Several earlier talks at this conference have emphasised the 
importance of a good knowledge of the low energy neutrino cross sections, 
the vast majority of which come from bubble chamber
experiments performed many years ago.  In particular they are needed for  Monte Carlo (MC) tuning and model
development. It is very important when adjusting the parameters of a MC programme to
fit new data, that agreement with  older data sets is not lost.

In the past comparisons with data have tended to be qualitative rather than 
rigorously quantitative.  There have been many cross section compilation plots at 
various workshops \cite{zeller} and there is a danger of data being transfered from plot to plot with the
possibility of inaccuracies creeping in and the original sources of the data
being lost. In addition since the data has previously been unavailable in a useful central form, these
plots have often been arrived at independently by each of the various speakers.

Our aim here is to produce the definitive quantitative database of validated low energy neutrino cross sections 
together with their associated statistical and systematic errors.
\section{THE DURHAM HEPDATA PROJECT}

\subsection{HEPDATA - introduction}

The development of this data resource is linked to the Durham HEP Database
project, HEPDATA, with which the author is involved.  A very brief description
of the HEPDATA project is given here specifically to demonstrate its relevance to the neutrino 
data resource.

HEPDATA is a PPARC(UK) funded project which has been in existence now for over
25 years.  Its three primary aims have remained essentially the same over this period,
namely: 
(a) To compile scattering data from all types of HEP reactions (cross sections,
polarisations, etc...)(b) To make the resulting compilations easily available to the whole community, and
(c) To engage in a programme of data commentary and evaluation - data reviews.

More recently other services such as the hosting of mirrors of the SLAC SPIRES
databases and the Berkeley PDG Review of Particle Physics web pages in the UK,
have been added to the HEPDATA operation. HEPDATA also provides a unique and comprehensive 
Parton Distribution Code server.
These are all accessible from the main HEPDATA home 
web page \footnote{http://durpdg.dur.ac.uk/hepdata/}.

\subsection{HEPDATA - database coverage}

The scope of the HEPDATA database covers cross sections from all types of particle physics reactions. It is
emphasised that it does not contain \lq\lq particle properties" which fall
into the domain of the RPP of the Berkeley PDG. It also not contain raw data such as
found on DSTs of experiments.  To appear in the database the data are generally in the final published
form. Ideally, to be most useful,  they should be fully corrected for acceptances and efficiencies and 
be model independent.
The database contains data from around 10000 publications dating from
the 1970s to the present day and is regularly updated.  The data are obtained
from journals and preprints and direct from the experiments 
especially when data appear only in graphical form in a publication. In the latter case the authors of the
paper are contacted to obtain the exact numerical values shown in the plot. It is
very important that this is done at the time of the publication as experience has
shown how difficult it is to obtain numerical values at a later date. Data are
rarely read from plots due to the difficulty in getting accurate representations
of the values, in particular the uncertainties.  
In its initial phase, over twenty years ago, the HEPDATA project mainly concentrated on 
cross sections above
2 GeV, therefore when attempting to retrieve a complete and reliable collection
of decades-old low energy cross section data, special care is needed to ensure 
completeness, and the latter methods need to be used more frequently.
Finally, verification of 
data entered into the database is always sought from the experimenters themselves.

\subsection{HEPDATA - data reviews}

A development of the work of  HEPDATA, which has direct
relevance to the subject of this talk, namely the creation of a neutrino 
data resource, is the involvement of the project in the production of Data Reviews. This 
fulfils the third stated aim of the project above, the programme of
data commentary and evaluation. To this end, over the past 15 years the 
project has produced a series of reviews on a variety of topical
and timely aspects of HEP data.  Many of these have been published in the IoP's
Journal of Physics G, and more recently have been made available 
on-line where they are kept up-to-date. In these reviews
all available data on the topic are presented in both numerical and
tabular form together with a short introduction usually written by an
expert in the field. There are also comparisons between data sets and, where
applicable, comparisons with the relevant theory curves.
Topics covered in Data Reviews to date include:
\begin{list}{$$}
{
\setlength{\parsep}{0.0cm}
\setlength{\itemsep}{0.0cm}
\setlength{\leftmargin}{0.0cm}
}
\item Structure Functions in DIS \cite{rev:f2}
\item Drell Yan Cross Sections \cite{rev:dy}
\item EE Correlations in $e^+e^-$ Interactions \cite{rev:eec}
\item Total Cross Sections in $e^+e^-$ Interactions \cite{rev:R}
\item Particle Production in 2$\gamma$ Interactions \cite{rev:2g}
\item Hadron Production in  $e^+e^-$ Interactions \cite{rev:ee}
\item Single $\gamma$ Production in Hadron Interactions \cite{rev:1g}
\end{list}
Fuelled in part by recent interest shown at the NuInt series of workshops,
the Low Energy Neutrino Data resource is envisaged to become the latest in the 
review series.

\section{THE NEW NEUTRINO DATABASE}
\subsection{Proposed Content}
We have identified the following conditions and limits for data to be 
included in the database:
\begin{list}{$\bullet$}
{
\setlength{\parsep}{0.0cm}
\setlength{\itemsep}{0.0cm}
\setlength{\leftmargin}{0.4cm}
}
\item Charged Current (CC) and Neutral Current (NC) data from both $\nu$ and $\bar{\nu}$ interactions.
\item E$_\nu \le$ 30 GeV.
\item Total and Differential cross sections from:
\begin{list}{-}
{
\setlength{\parsep}{0.0cm}
\setlength{\itemsep}{0.0cm}
\setlength{\leftmargin}{0.0cm}
}
\item Inclusive reactions ($\nu_\mu N \rightarrow \mu^- X$, etc...).
\item Quasi-Elastic (QE) reactions ($\nu_\mu n \rightarrow \mu^-  p$, etc...).
\item Single $\pi$ exclusive reactions (both incoherent and coherent).
\item Two $\pi$ exclusive reactions.
\end{list}
\item Fluxes (absolute normalisation and shape.)
\end{list}
Knowledge of the intensity and shape of the incident (anti)neutrino
flux and their uncertainties is very important in producing an
accurate and reliable database of cross sections. It is
particularly important for interpreting and using differential
cross section distributions from the experiments (which are 
included in the database). Significant effort will therefore
be made to ascertain these for the individual data sets.

\begin{table}[t]
\caption{Summary of data from the different experiments}
\label{table:exp}
\newcommand{\m}{\hphantom{$-$}}
\newcommand{\cc}[1]{\multicolumn{1}{c}{#1}}
\renewcommand{\tabcolsep}{2pt} 
\begin{tabular}{llccc}
\hline
Accelerator  & Facility & Incl. & QE & Excl. $\pi$(s) \\
\hline
CERN-PS   &  Gargamelle   & \cite{ggm1,ggm2,ggm3} & \cite{ggm5,ggm6,ggm7} & \cite{ggm8,ggm9,ggm10} \\
CERN-SPS  &  Gargamelle   & \cite{ggm4} &- &  -\\
          &  BEBC         & \cite{bebc1,bebc2,bebc3,bebc4} & \cite{bebc5} & \cite{bebc5,bebc6,bebc7,bebc8,bebc9} \\
          &  CHARM        & \cite{charm1,charm2} &	-  & \cite{charm3,charm4}\\
          &  CDHS         & \cite{cdhs1} &	-  &	-     \\
FNAL  &  15ft BC    	  & \cite{fnal1,fnal2,fnal3} & \cite{fnal4} &\cite{fnal5,fnal6,fnal7}\\
          &  CCFRR        & \cite{ccfrr1} &	  -    & -	 \\
          &  CCFR         & \cite{ccfr1,ccfr2} &	-      & -	 \\
ANL   &  12ft BC    	  & \cite{anl1,anl2} & \cite{anl3,anl4} & \cite{anl2,anl5,anl6,anl7} \\
BNL&  7ft BC     	  & \cite{bnl1,bnl2,bnl3} & \cite{bnl2,bnl4} & \cite{bnl5} \\
Serpukhov &  SKAT         & \cite{skat1} &  \cite{skat2} &  \cite{skat3,skat4,skat5}\\
          &  ITEP         & \cite{itep1,itep2} & \cite{itep3,itep4,itep5} &    -      \\
          &  JINR         & \cite{jinr1} &   -       &   -       \\
LANL&  LSND               & \cite{lsnd1} &   -       &     -     \\
\hline
\end{tabular}\\
\vspace*{-1.0cm}
\end{table}

\subsection{Locating the data}

In Table \ref{table:exp} we list the accelerators and
experiments/facilities where these measurements have been made,
indicating the type of data available together with the relevant publication references.  The data sources range from
the very early Gargamelle experiments at CERN in the 1970's to the more
recent experiments at CERN, FNAL and Serpukhov.  There is also a very 
recent inclusive cross section measurement from LSND at LANL.

The relevant data has been collected from a variety of sources.  
Whenever possible we have obtained the numbers from tables or
text in the publications, or used data which was originally 
supplied by the authors of the papers. This ensures that we have
exact and accurate data, including their uncertainties. We have also obtained 
numbers from private data collections.  
As a last resort we have read data from plots in the papers.
The accuracy and validity of the latter sources of data points is obviously more
uncertain  and it is our
intention to have these verified by physicists who were
members of the original experimental collaborations.

\begin{figure*}[t]
\epsfig{file=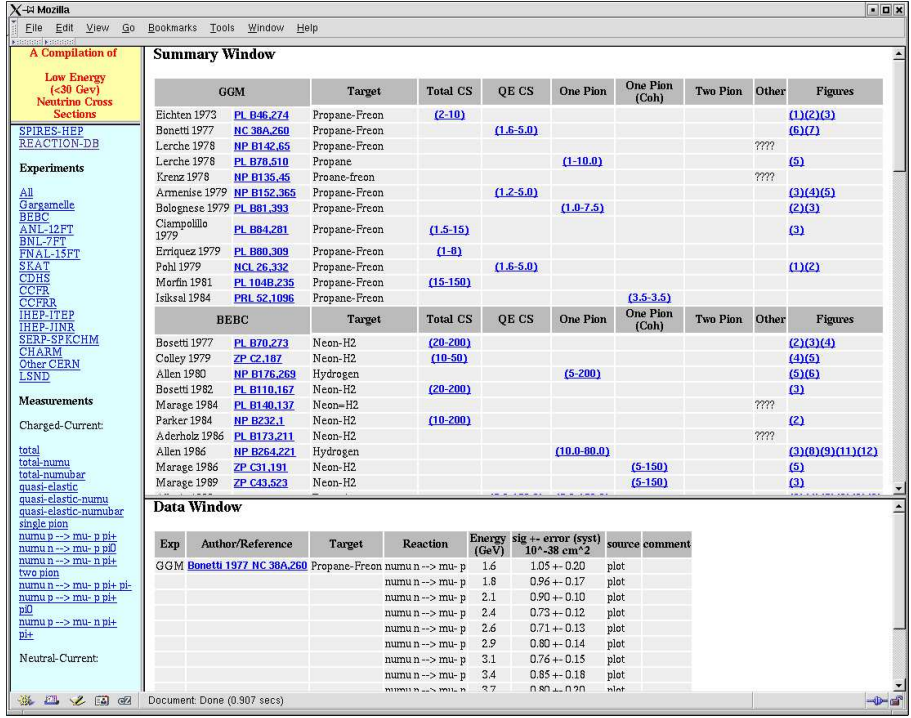,width=0.75\textwidth}
\vspace*{-0.75cm}
\caption{The main home web page of the Neutrino Data Resource\hfill\break\hspace*{1.5cm}( http://durpdg.dur.ac.uk/hepdata/online/neutrino/)}
\label{fig:home}
\vspace*{-0.25cm}
\end{figure*}

\subsection{The MySQL database}

A simple relational database using the MySQL
database management system has been constructed
to store the information and data values.
The database has only two \lq\lq database tables".
The first contains information relevant to the 
whole paper including
the target material, beam energy range and the source of the data, as well
as more general bibliographic information. The
second table contains the actual data points and a
field for general comments.  The tables are linked
using the IRN number from the SLAC/SPIRES database
entry for the paper. This is a unique and permanent
number assigned to every paper by the SLAC library
group who manage the SPIRES databases. It is a very
useful reference number used also in the HEPDATA databases.

\subsection{The Neutrino Web Site}

\begin{figure*}[t]
\epsfig{file=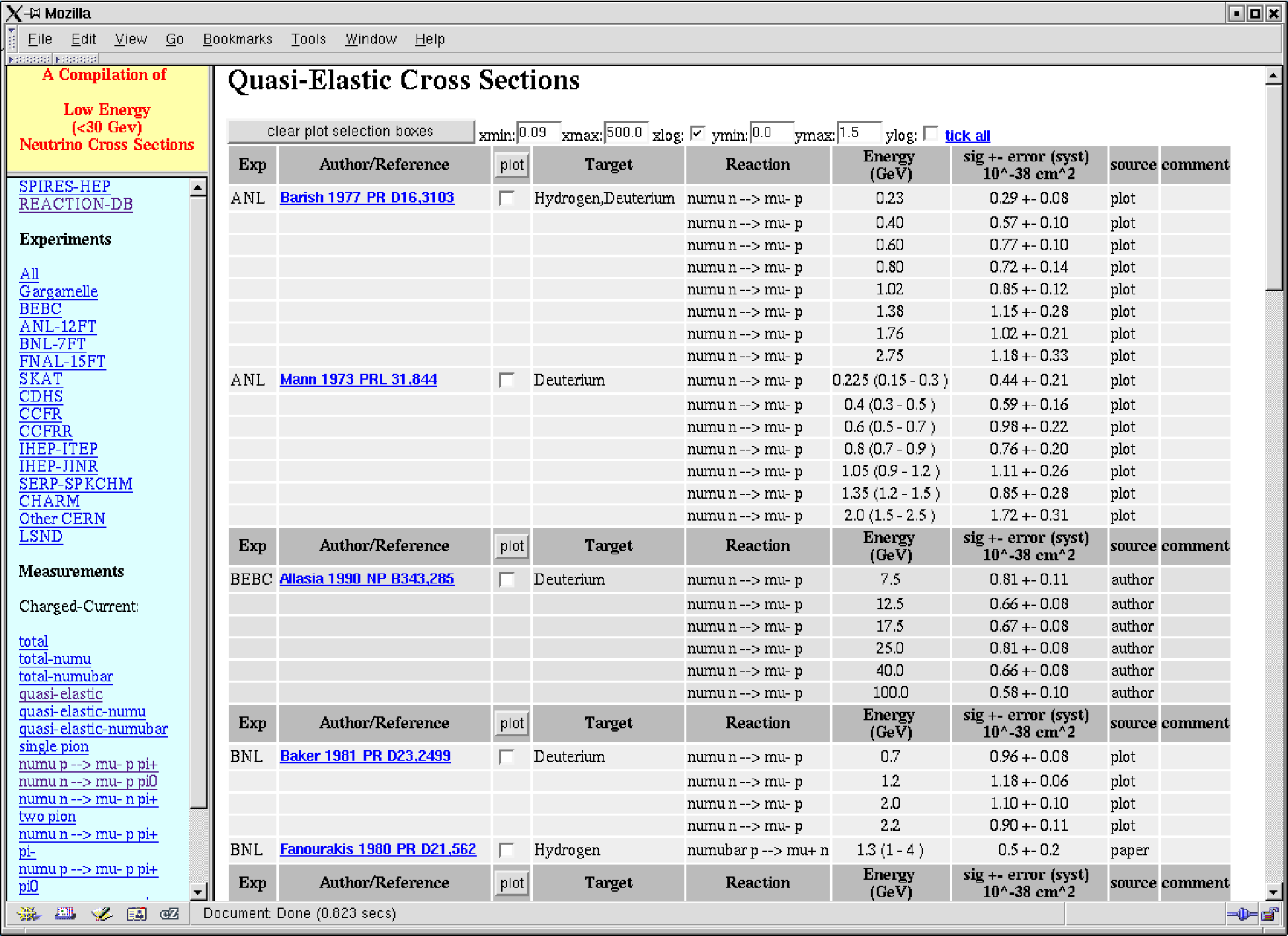,width=0.75\textwidth}
\vspace*{-0.75cm}
\caption{The Quasi-Elastic web page of the Neutrino Data Resource}
\label{fig:qe}
\vspace*{-0.25cm}
\end{figure*}

From the neutrino database a prototype web site is automatically generated
giving user access to the data and information. Figure \ref{fig:home} shows the 
main home page of this site.
The page is basically divided into two main frames plus a navigation
frame at the side on the left.  This navigation frame, as its name suggests,
remains available throughout all other pages of the web site.   

From the upper part of this navigation space the user can
select data from specific, or all (the initial default), experiments  with the
relevant information appearing in the upper right hand frame, the Summary Window.  Each row in this
table represents a publication by the selected experiment with columns for
the first author on the paper, the publication reference, the target material and the type of measurement (inclusive,
QE or pion(s)).  A final column gives links to the relevant plots from the
paper. 

The publication reference provides a link to the SLAC/SPIRES database record using
the IRN number.  This gives the user quick access in most cases to the paper itself and other
bibliographic information. The columns for the types of measurements contain the 
neutrino beam energy ranges which are themselves selectable links to the actual 
numerical data. Selecting one of these links will cause the relevant data to be displayed 
in the lower right hand frame, the Data Window. As an illustration, in Figure \ref{fig:home} 
the result of selecting the
QE Gargamelle data from the Bonetti 1977 paper is shown in the Data Window.      
This appears as a table with a row for each measured data point.  The table columns for the 
experiment name, the author/reference (again
linked to the SLAC/SPIRES database), the target material, the reaction, the beam energy,
the measured cross section, the source and optional comments.  These columns are somewhat
self explanatory. The energy is given as the mean value and where available 
the range is also shown.  The cross section values are shown with separate systematic 
errors where they are given in the papers.  The source column indicates how the data was obtained:
read from a {\em plot}, from numbers in the {\em paper} or directly from the {\em author}s. 
As we are still in the process of fully vetting the data, the method of their aquisition should serve
as an indicator of the robustness of the numerical values. 

\begin{figure}[h]
\epsfig{file=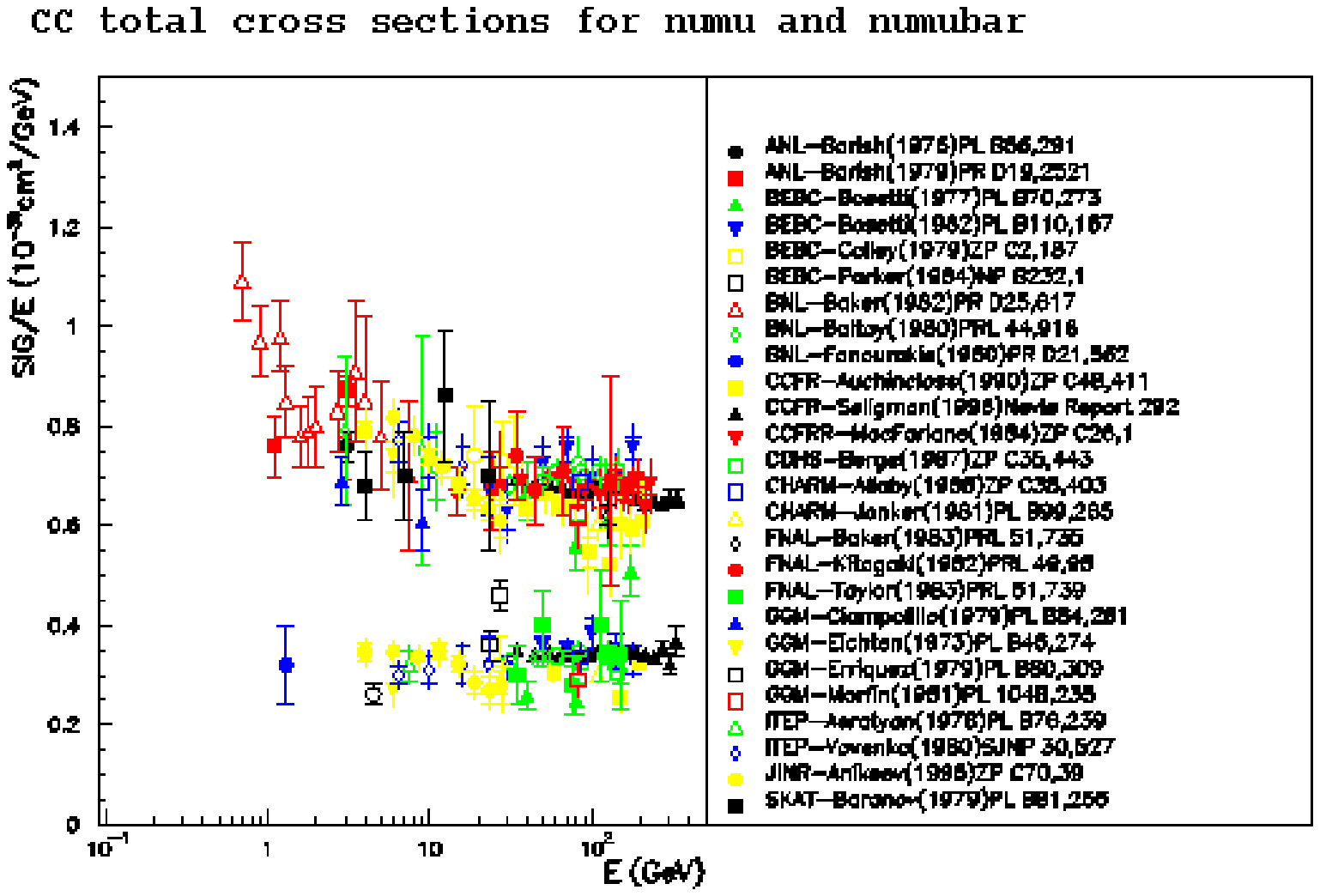,width=0.45\textwidth}
\vspace*{-0.75cm}
\caption{Plot of all CC total inclusive data generated from the database}
\label{figure:plot}
\vspace*{-0.75cm}
\end{figure}

The lower half of the navigation space has links to the data displayed in measurement types
such as total inclusive, quasi-elastic and single and double pion exclusive, cross sections.
Figure \ref{fig:qe} shows the result of selecting the QE cross sections.  The display changes to a single
panel but the columns are similar to those of the Data Window display from the experiments selection.
In addition there is the facility on this page to make "on-the-fly" plots of one or more data 
sets.  To do this the user simply ticks the relevant boxes in the \lq\lq plot" column and then
selects any of the plot buttons.  At the top the facility to choose the range and type of the  
scales of the axes of the plots. Figure \ref{figure:plot} shows and example of a plot of the
total CC inclusive cross sections generated from the database in this way.

\section{Summary}
A new collaborative project has been described to produce a definitive quantitative database
of low energy neutrino cross sections, initially covering total inclusive, QE and 
single and double pion exclusive, processes.  It is the intention to develop the project 
to include the results of studies of the systematic uncertainties
and flux normalisations reported in the papers and in private communications. In achieving the latter
we intend to enlist the help of  members of the community who worked on the early neutrino
experiments and who are still active in the field to provide the
best possible verified data.

An embryonic database and web page have been constructed to provide the user with advance and preliminary access to the database.
Other methods of accessing
and using the data, such as directly from the database, will be developed in the future. One immediate user 
of the new database
will be the Neutrino Monte Carlo Validation Tool, being developed by  Costas Andreopoulos,
where the output from various MC programs can be \lq\lq tuned" against existing
data.   This will be described in the next talk \cite{costas}.

\section{Acknowledgements}
The author wishes to than all the members of this collaborative effort for their input and
help.  He also wishes to acknowledge the PPARC(UK) for their continued support of the 
HEPDATA project under grant number PPA/G/O/2001/00624.

\end{document}